# Linking Social Networking Sites to Scholarly Information Portals by ScholarLib

**Peter Mutschke**
GESIS – Leibniz Institute for the
Social Sciences
Unter Sachsenhausen 6-8
D-50667 Köln, Germany
peter.mutschke@gesis.org

**Mark Thamm**
GESIS – Leibniz Institute for the
Social Sciences
Unter Sachsenhausen 6-8
D-50667 Köln, Germany
mark.thamm@gesis.org

## Abstract
Online Social Networks usually provide no or limited way to access scholarly information provided by Digital Libraries (DLs) in order to share and discuss scholarly content with other online community members. The paper addresses the potentials of Social Networking Sites (SNSs) for science and proposes initial use cases as well as a basic bi-directional model called ScholarLib for linking SNSs to scholarly DLs. The major aim of ScholarLib is mutually bringing together the content of SNSs and DLs in order to enrich SNSs by scholarly content retrieved from DLs and, vice versa, to enhance searching at DL side by social information provided by SNSs.

## ACM Classification Keywords
H.3.5 [On-line Information Services]: Web-based services; H.3.7 [Digital Libraries]; H.5 [Information interfaces and presentation]: Web-based interaction

**Introduction**

The Social Web represents a fundamental change of the roles users play in the Internet [1,3,4,5,6]. With the help of collaborative features the Social Web provides novel models of knowledge creation and exchange that enhance user-generated content and collective intelligence abolishing a strict distinction between producer and consumer of information. Social Networking Sites (SNSs) like Facebook[1] or LinkedIn[2] moreover offer innovative frameworks for community building and knowledge dissemination. The particular benefits of SNSs, furthermore, are the viral effects that are taken place on those platforms. Due to the inherent social dynamic of SNSs and the transitivity of network relationships content shared among members may spread throughout the network.

These features suggest new models of collaborative knowledge creation and exchange also for science. However, there is still a gap between the traditional world of Web1.0 fashioned scholarly information systems such as Digital Libraries (DLs) and the new world of the Social Web. Despite the fact that many SNSs nowadays are open to applications that can be added to the platform for collaborative usage (by providing APIs such as OpenSocial[3]) many third-party applications on SNSs, in particular on Facebook, are for non-professional usage. Up to now there are only a few professional or scholarly platforms that are integrated with SNSs. First examples are the online library WorldCat[4] that can be accessed via a search plugin on Facebook[5], and the e-print archiv arXiv[6] that enables Facebook users to comment arXiv articles and dynamically list arXiv articles on their Facebook page[7].

To the best of our knowledge, SNSs usually provide no or limited way to access scientific information provided by scholarly DLs in order to share and discuss scholarly content with other members of online communities. The paper therefore addresses the potentials of SNSs for science and proposes initial use cases as well as a basic bi-directional model for linking SNSs to scholarly DLs, and vice versa.

**Potentials of Web 2.0 for Science**

Handling knowledge in an efficient way is a crucial success factor for science. Keeping this in mind, Web2.0 platforms like SNSs provide a high potential for new innovative forms of scholarly knowledge production and dissemination [8]. Four major benefits can be identified for using SNSs in science: (1) SNSs provide platforms for getting digitally networked with colleagues having similar research interests. (2) As tools for posting novel information to a network of contacts SNSs provide the chance to publish and distribute research results before being published via classical media [6]. (3) Due to the virality of SNSs, information entered to an SNS might be spread throughout the whole social graph of the SNS, to the degree to which contacts (respectively contacts of contacts and so on) share information with others such that scholarly information can be easily disseminated to a wider community. (4) Corresponding to (2) SNSs provide the chance to find relevant scholarly content

---

[1] http://www.facebook.com/
[2] http://www.linkedin.com/
[3] http://docs.opensocial.org/display/OS/Home
[4] http://www.worldcat.org

[5] http://apps.facebook.com/worldcat/
[6] http://arxiv.org
[7] http://apps.facebook.com/myarxiv

posted to a SNS much earlier than by classical channels such as DLs [6].

SNS platforms therefore already started to enter scientific working. There are first successful examples of scholarly SNSs (Mendeley[8], ResearchGate[9], Bibsonomy[10]) that provide the maintenance and exchange of scientific documents, social tagging and networking. Recent studies [3,5,8,10], such as the CIBER report on the role of social media for the research workflow, actually confirm that SNSs are seen as useful tools for "identifying research opportunities", "research collaboration", "reviewing the literature", "collecting research data" and "disseminating research findings" [3].

Currently, however, SNSs are still far away from being well established as suitable tools for science [2]. A major reason for this is to be seen in the fact that up to now SNSs are not well coupled with classical DLs such that the user is not enabled to enrich SNS content with research information provided by scholarly information portals. The latter, however, would enhance the scholarly alignment of SNSs significantly - a crucial condition for the acceptance of SNSs as platforms for scientific communication. From the perspective of scholarly DL providers, on the other hand, linking DLs with SNSs might increase their usage immensely, due to growing use of the Web 2.0 also by scientists [1,5]. Some authors moreover argue that DLs risk losing a majority of the scholarly information flow and, as a possible result, their foundation as information intermediaries if the chances of Web 2.0 for science are not used by scholarly DL providers [9].

It seems that a symbiotic relationship between the two "worlds" can be identified here: to the same extent to which an integration of Web 2.0 in classical information systems plays a crucial role for the justification of traditional information intermediaries, the enrichment of SNSs by scholarly content will be a major factor of being accepted in science. This suggests an integration of scholarly DLs with SNSs [5]. A coupling of SNSs with scholarly DLs, however, should work in both directions, i.e. by building bridges between the two that can be entered from both directions - ensuring that scholarly information can be accessed via SNSs and social information related to scholarly issues can be addressed by scholarly portals.

**ScholarLib**
The aim of the project presented, referred to as *ScholarLib* in the following, is therefore to develop a generalized framework for bi-directional linking SNSs to scholarly DLs - in order to make scholarly information provided by DLs accessible at SNSs, and vice versa, to enrich scholarly content at DL side by social information provided by SNSs [11]. By mutually bringing together the content of SNSs and DLs it is expected that ScholarLib may generate more virality on scholarly communication and therefore may contribute to community building and knowledge exchange in science. The main use cases for such a coupling, currently under development in the project, are (where →indicates access direction and >> posting direction):

---

[8] http://www.mendeley.com
[9] http://www.researchgate.net
[10] http://www.bibsonomy.org

*Searching scholarly information (SNS→DL)*
The basic use case of ScholarLib is to enable a SNS user to perform a search in a scholarly DL via a particular search plugin installed at the SNS. The DL returns a result set for the query which is presented at the applications canvas page within the SNS (see Figure 1). The resulting data from the DL contains the scholarly items retrieved together with annotations from other users of ScholarLib that had already been added to items in the result set. An interesting variant of this use case is to apply the search as an alerting service in due consideration of profile data such as research interests. This could be furthermore enhanced by enriching the user's profile data by controlled scholarly vocabulary provided by search term recommendation services of the DL [7].

*Posting social information (SNS>>DL)*
Once the retrieved scholarly items are stored at SNS side, e.g. in a personal library, the SNS user is now allowed to annotate items by storing, rating, commenting, tagging or linking items (see Figure 1) such that scholarly information, usually described and indexed by information specialists at DLs, is now enriched by additional information from the perspective of the user. Moreover the SNS user is able to share information with others such that scholarly information may increase its "social impact" due to the degree to which it is spread throughout the social graph of the SNS. Social information on the items retrieved is finally passed back to the DL such that it can be used for indexing and ranking in order to enhance retrieval quality at DL side.

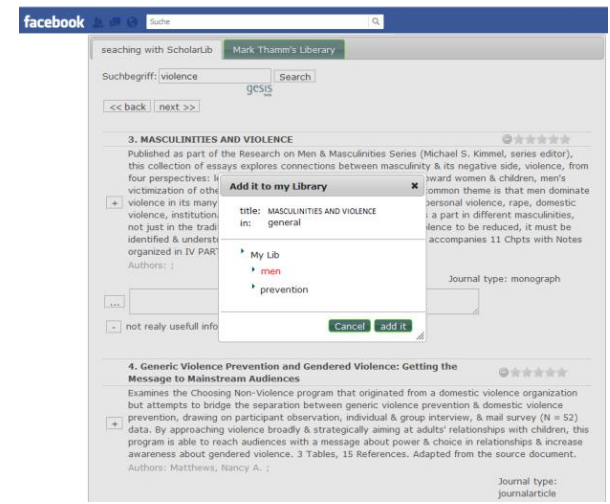

**Figure 1.** First prototype of ScholarLib for Facebook: A Facebook user performs a search for the term 'violence' and is then provided by hits from the Social Science portal sowiport[11]. The user tags selected items by adding them to a particular folder (here 'men') of the user's local library.

*Searching social information (DL→SNS)*
The inversion of the first use case is to allow a DL user to search for social information provided by an SNS. This may include profile pages of actors with similar research interests, discussions on SNS side that match with the user's query, as well as annotations and ratings of scholarly items related to the user's topic (see first use case). ScholarLib retrieves this information from the SNS and adds it to the result set of the search performed at DL side.

---

[11] http://www.sowiport.de

*Posting scholarly information (DL>>SNS)*
The inversion of the second use case is to allow a DL user to post scholarly items identified as relevant (e.g. publications in a result set for a query) to the own SNS page or to members of the own network at the SNS having similar research interests. Corresponding to the alerting service seen for the first use case posting of selected items could be also performed pro-actively.

## Architecture

The technical main issues of ScholarLib are twofold: (1) the development of an open communication channel between SNSs and DLs by providing a protocol for mutual information exchange, and (2) a technical framework that handles scholarly as well as social information as addressable objects.

Due to (1) ScholarLib is divided into three loosely coupled parts (see Figure 2): the SNS connector layer, the core services layer (the internal part of ScholarLib), and the DL connector layer. The main design objective here is to make further integrations of other platforms – either DLs or SNSs - as simple as possible. The DB connectors act as the interface to the DLs. DLs to be connected need to provide an http-accessible search interface which delivers scholarly items in a specific format (currently a subset of Dublin Core). By registering the interface's URL the DL is connected to ScholarLib. The SNS connectors implement APIs to the SNSs to be connected, currently OpenSocial[12] and OpenGraph[13]. By installing the connector as an application within the SNS's runtime environment the SNS is connected to ScholarLib.

---

[12] http://docs.opensocial.org/display/OS/Home
[13] http://developers.facebook.com/docs/opengraph/

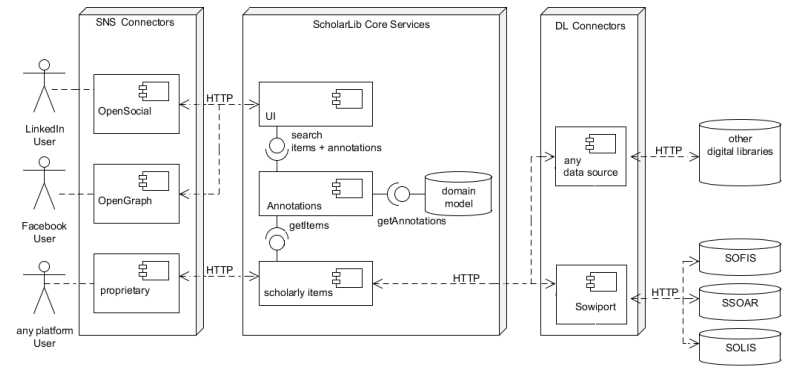

**Figure 2.** Software architecture of ScholarLib, consisting of three loosely coupled parts, the SNS connector layer, the core services layer, and the DL connector layer.

The main task of ScholarLib's core, which resides between the two connector layers, is passing search queries and result sets from one system to the other. Due to (2) all use cases of ScholarLib require the maintenance of persistent relations between SNS users and scholarly items on which social activities take place, such as annotations and sharings. The domain model of ScholarLib (see Figure 3) therefore consists of the two entity types SNUser (SNS user) and SItem (any kind of scholarly item) as well as several relationships representing social activities of SNUsers on SItems (Comment, Rating, Forward, Library). Creating persistent relations between scholarly actors on SNSs and scholarly items retrieved from DLs is seen as the internal key functionality of ScholarLib since on the basis of such relations a number of enhanced services are conceivable (e.g. recommender services). Thus, retrieving and storing data and relations is done within a persistence layer which is strictly separated from the UI layer.

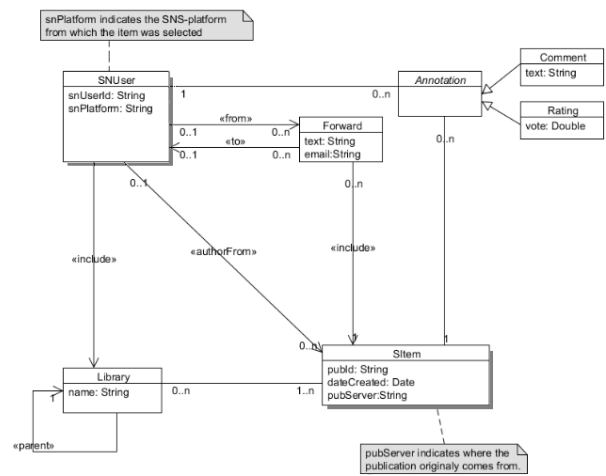

**Figure 3.** Domain model of ScholarLib: main entity types are SNUser (SNS user) and SItem (scholarly item) that are tied to each other by social activities of SNUsers on SItems.

## Future Work and Challenges

ScholarLib will be prototypically implemented for the field of Social Sciences. First prototypes are currently under development for the first use case using sowiport as DL and Facebook as well as the German scholarly Web 2.0 platform iversity[14] as SNSs. In a further step we intend to apply ScholarLib also to DLs from other scientific disciplines. We furthermore aim at setting up ScholarLib as a platform to investigate how social information retrieval is affected by the virality of SNSs (and vice versa), which is seen as a major research challenge. A particular focus of the project will therefore be to trace the flow of scholarly information throughout the social graph of SNSs in order to use such information for enhanced search services.

---

[14] http://www.iversity.org/

**References**

[1] Busemann, K., Gscheidle, C. Web 2.0: Communitys bei jungen Nutzern beliebt. Ergebnisse der ARD/ZDF-Onlinestudie 2009. *Media Perspektiven*, vol. 7, 2009, 356-364.

[2] Büffel, S., Pleil, T., Schmalz, J.S. Net-wiki, pr-wiki, kowiki – Erfahrungen mit kollaborativer Wissensproduktion in Forschung und Lehre. *kommunikation@gesellschaft*, vol. 8, 2007.

[3] CIBER report *Social media and research workflow.* University College London, Emerald Group Publishing Ltd, 2010.

[4] Ebersbach, A., Glaser, M., Heigl, R. *Social Web*. UVK Verlagsgesellschaft mbH, Konstanz, 2008.

[5] Evans, B.M., Kairam, S., Pirolli, P. Do your friends make you smarter?: An analysis of social strategies in online information seeking. *Information Processing and Management*, vol. 46, 2010, 679-692.

[6] Koch, D. *Onlinestudie: Wissenschaftliches Arbeiten im Web 2.0*. urn:nbn:de:0009-5-18425, 2009.

[7] Mutschke, P., Mayr, P., Schaer, P., Sure, Y. Science Models as Value-Added Services for Scholarly Information Systems. *Scientometrics*, 89 (1), 2011, 349-364.

[8] Nentwich, M., König, R. *Wissenschaft und Social Network Sites*. ITA-Projektbericht Nr. A-52-5, 2011.

[9] Schindler, C., Rittberger, M. Herausforderungen für die Gestaltung von wissenschaftlichen Informations-infrastrukturen durch Web 2.0. *Information - Wissenschaft & Praxis*, vol. 60 (4), 2009, 215-224.

[10] Shapira, B., Zabar, B. Personalized Search: Integrating Collaboration and Social Networks. *JASIST*, 62(1), 2011, 146-160.

[11] Thamm, M., Wandhöfer, T., Mutschke, P. ScholarLib: Ein Framework zur Kopplung von Sozialen Netzwerken mit wissenschaftlichen Fachportalen. In *Proc. 64. DGI Annual Meeting*, 2012.